# The comparison of Higuchi's fractal dimension and Sample Entropy analysis of sEMG: effects of muscle contraction intensity and TMS

Milena B. Čukić[1,*], PhD, Mirjana M. Platiša[2], PhD, Aleksandar Kalauzi[3], PhD, Joji Oommen[4], MS, Miloš R. Ljubisavljević[4], MD, PhD

[1]Department for Physiology with Biophysics, School of Biology, University of Belgrade, Belgrade, Serbia

[2]Institute of Biophysics, School of Medicine, University of Belgrade, Belgrade, Serbia

[3]Department for Life Sciences, Institute for Multidisciplinary Research, University of Belgrade, Belgrade, Serbia

[4]Department for Physiology, College of Medicine and Health Sciences, UAE University, Al Ain, UAE

# Abstract

The aim of the study was to examine how the complexity of surface electromyogram (sEMG) signal, estimated by Higuchi's fractal dimension (HFD) and Sample Entropy (SampEn), change depending on muscle contraction intensity and external perturbation of the corticospinal activity during muscle contraction induced by single-pulse Transcranial Magnetic Stimulation (spTMS). HFD and SampEn were computed from sEMG signal recorded at three various levels of voluntary contraction before and after spTMS. After spTMS, both HFD and SampEn decreased at medium compared to the mild contraction. SampEn increased, while HFD did not change significantly at strong compared to medium contraction. spTMS significantly decreased both parameters at all contraction levels. When same parameters were computed from the mathematically generated sine-wave calibration curves, the results show that SampEn has better accuracy at lower (0-40 Hz) and HFD at higher (60-120 Hz) frequencies. Changes in the sEMG complexity associated with increased muscle contraction intensity cannot be accurately depicted by a single complexity measure. Examination of sEMG should entail both SampEn and HFD as they provide complementary information about different frequency components of sEMG. Further studies are needed to explain the implication of changes in nonlinear parameters and their relation to underlying sEMG physiological processes.

# Introduction

Surface EMG (sEMG) is a record of electrical activity of underlying muscle fibers. It is a complex, nonlinear and non-stationary signal, influenced by factors like neuron discharge rates, recruitment patterns of motor units, muscle architecture, as well as various other factors (Nieminen and Takala, 1996). The analysis of sEMG has extended beyond the traditional diagnostic applications to also include applications in diverse areas such as biomedical, prosthesis or rehabilitation devices, human machine interfaces, and more. Traditionally the sEMG analysis, particularly in the medical field, is dominated by spectral or amplitude measures. Nevertheless, it was repeatedly shown that nonlinear analyzes of sEMG can provide additional information on the underlying motor strategies (Del Santo et al., 2007; Farina et al., 2002), hidden rhythms (Filligoi and Felici, 1999), fatigue (Ikegawa et al., 2000) as well as detection of pathological changes in the system (Meigal et al., 2012, 2009). Furthermore, several studies showed that nonlinear methods are potentially more sensitive than classical sEMG analysis methods, being able to capture very subtle changes in the signal under study. For example, fractal analysis (Mesin et al., 2009; Ravier et al., 2005) was used to show the relationship between muscle force changes during contraction and the complexity of the sEMG signal (Gitter and Czerniecki, 1995; Gupta et al., 1997) and to detect the lowest level of the voluntary activation of the muscle under study (Arjunan and Kumar, 2010). Nonlinear methods have also proven useful in distinguishing between sEMG of patients with Parkinson's disease and healthy controls (Meigal et al., 2009), potentially providing useful biomarkers of system dysfunction in aging and disease.

Transcranial Magnetic Stimulation (TMS) is a noninvasive method used to stimulate brain cortex. When applied over motor cortex it induces synchronous activation of corticospinal neurons, which in sEMG evokes characteristic activation, termed Motor Evoked Potential (MEP). After MEP a temporary silence of sEMG activity called silent period (SP) occur. It was accepted that the effects of this single-pulse TMS (spTMS) of the motor cortex do not induce lasting changes in cortical activity beyond those immediate ones. However, our recent study showed that the complexity of the sEMG signal after single-pulse TMS decreases suggesting that the overall corticospinal activity after spTMS becomes less complex (Cukic et al., 2013). The study used Higuchi's fractal dimension (HFD) (Higuchi, 1988) which serves as a measure of signal complexity. Fractal dimension (FD) refers to a non-integer or fractional dimension of a geometric object; it could be used for phase space approach to estimate the FD of an attractor in the state-space domain. Applications of HFD within this framework pertain primarily to time domain since the signal itself is considered as a geometric figure (Kalauzi et al., 2012). However, it has been suggested that nonlinear measures can overestimate or underestimate subtle changes in the signal complexity with different algorithms yielding different results for a single nonlinear measure (Ferenets et al., 2006; Goldberger et al., 2002; Mesin et al., 2009).

Sample entropy (SampEn), another nonlinear method, was first introduced as a 'regularity' statistics (Richman and Moorman, 2000) rather than a direct index of physiological complexity. SampEn quantifies the probability that sequences of patterns in a dataset that are initially closely related remain close on the next incremental comparison, within a specified tolerance. Thus, SampEn appears to be a potentially useful method to unravel the complexity of the sEMG signal and its relation to underlying physiological processes (Cashaback et al., 2013; Lake et al., 2002; Molina-Picó et al., 2011; Ruonala et al., 2014). Very few studies explored the use of SampEn to analyze EMG signal. For example, the complexity of biceps sEMG was shown to exhibit a complex relation to contraction intensity (Cashaback et al., 2013). Muscle fatigue decreased the complexity of sEMG towards the end of fatiguing-exhausting contraction (Cashaback et al., 2013), as well as the complexity of submaximal and maximal voluntary contraction (Pethick et al., 2015). Furthermore, various nonlinear complexity parameters were shown to be significantly less variable in differentiating non-fatiguing and fatiguing muscle contraction (Karthick et al., 2014), making them potentially useful for automated analysis of neuromuscular activity in normal and pathological conditions. Nevertheless, the mechanisms influencing sEMG complexity are still poorly understood. Therefore, to further explore the use of non-linear methods in sEMG analysis, potentially broadening their clinical applications, in this study, we compared SampEn with HFD using the same interference spTMS paradigm. We examined the changes in sEMG complexity by SampEn during voluntary contraction of different intensities before and after application of spTMS. Finally, to elucidate the difference in obtained results we compared the results of SampEn and HFD analysis of theoretical mathematically constructed calibration curves (Kalauzi et al., 2012).

## Materials and Methods

*Participants*

Surface electromyogram (EMG) was recorded from the first dorsal interosseous muscle (FDI) of each participant. The sample comprised of ten participants, five women, five men (age 22-48 +/- 6.7 years). All participants were healthy volunteers, without a prior history of neuromuscular disorders.They were all righthanded, according to Edinburgh Inventory (Oldfield, 1971). Participants gave their written informed consent before the experimental procedure. The study was approved by the Al Ain Medical District Human Research Ethics Committee (Protocol

No. 12/44) and performed in accordance with the ethical standards laid down in the Declaration of Helsinki.

*Experimental protocol and conditions*

Participants were comfortably seated in an armchair, resting the right hand on a handhold. They were asked to exert voluntary FDI contraction by abducting the index finger against elastic resistant adjusted for each subject and contraction level. The intensity of contraction was provisionally expressed as a percentage of maximal voluntary contraction (MVC) and scaled as follows: mild (10-20%), medium (20-40%) and strong (40-70%). The contraction intensity was randomly varied. Subjects received up to 20 TMS pulses at each intensity, out of which 15 were used for the analysis. Sufficient time was given after each muscle contraction-spTMS trial, with longer time availed after strong contractions to avoid development of muscle fatigue. Subsequently, individual trials were grouped based on contraction intensity for further analysis.

*Transcranial Magnetic Stimulation and sEMG*

Transcranial magnetic stimulation (MagPro, R100, MagVenture, Denmark) was performed using with a figure-of-eight coil optimally positioned over the left hemisphere to evoke MEPs in the right FDI muscle (45 degrees to the central line). The optimal stimulation spot was marked with a semi-permanent marker to allow maintenance of stable stimulation coil position during the experiment. The resting motor threshold (RMT) was determined to the nearest 1% of the stimulator output and was the minimum intensity required to evoke MEPs of 50 µV in five out of ten consecutive trials (Rossini et al., 1994). The mean RMT for all subjects was $46.3 \pm 8.6\%$ of the maximal stimulator output. Subsequently, spTMS stimulus intensity was set at 1.3 above resting motor threshold.

Ag-AgCl electrodes were used to record the surface EMG from the right FDI muscle (electrode diameter 9 mm). The raw EMG signal was amplified and filtered with the band-pass filter in the range of 20 Hz – 1 KHz (CED 1902 isolated pre-amplifier, Cambridge Electronic Design, UK). Each recording was 8-10 s long with spTMS delivered approximately 4 s after the contraction onset. sEMG signals were digitized with the sampling rate of 1 KHz (CED 1401, Cambridge Electronic Design, Cambridge, UK) and stored for further off-line analysis. During the experiment, a Root Mean Square of the sEMG signal was computed and together with an audio signal shown to subjects.

*Data analysis*

From the recorded sEMG two epochs were selected for analysis: PRE TMS activity starting from the onset of the stable contraction up to approximately 10ms before TMS artifact (see Figure 1) and POST TMS activity, starting from the onset of uninterrupted sEMG (ignoring occasional mid-SP EMG bursts) after the SP. The beginning of post-epoch was estimated visually by the same examiner, a method shown to yield consistent detection not different from automated routines (Julkunen et al., 2013). Irrespective of individual variations in SP duration and time needed to develop stable contraction before spTMS, the length of both epochs used for analysis was set to 3.5 s. The epoch length did not differ more than 5 ms, which could not influence the analysis. In total, approximately 900 individual epochs (15 PRE and 15 POST-TMS, from 10 subjects, and three levels of contraction) were analyzed, and the output was used for the statistical analysis.

*Higuchi Fractal Dimension (HFD)*

The fractal dimension of sEMG was calculated by using Higuchi's algorithm (Higuchi, 1988), as a measure of signal complexity in the time domain. Higuchi proposed an algorithm for the estimation of fractal dimension directly in the time domain without reconstructing the strange attractor. This method gives a reasonable estimate of the fractal dimension even in the case of short signal segments and is computationally fast. EMG was analyzed in time, as a sequence of samples x(1), x(2),..., x(N), and k new self-similar time series $X_k^m$ were constructed as:

$$X_k^m : x(m), x(m+k), x(m+2k)...., x(m+\text{int}[(N-m)/k]k) \quad (1)$$

for m = 1, 2, ..., k where m is the initial time; k = 2, ...., $k_{max}$, where k is the time interval. According to previous studies which dealt with the application of Higuchi's algorithm with varying $k_{max}$ (Spasic et al., 2005), for this type of signals, the best option is $k_{max}$= 8. Int[r] is the integer part of a real number r. The length of every $L_m(k)$ was calculated for each time series or curves $X_k^m$ as:

$$L_m(k) = \frac{1}{k}\left[\left(\sum_{i=1}^{\text{int}[\frac{N-m}{k}]} |x(m+ik) - x(m+(i-1)k)|\right)\frac{N-1}{\text{int}[\frac{N-m}{k}]k}\right] \quad (2)$$

$L_m(k)$ has to be averaged for all m, therefore forming an average value of a curve length L(k) for each k=2,..., $k_{max}$

$$L(k) = \frac{\sum_{m=1}^{k} L_m(k)}{k} \quad (3)$$

Fractal dimension was evaluated as the slope of the best-fit form of ln(L(k)) vs. ln(1/k):

$$FD = \ln(L(k))/\ln(1/k). \quad (4)$$

Fractal dimensions were calculated separately for each epoch (PRE and POST TMS) using Matlab 7.0, using the computation reported earlier by Kalauzi et al. (Spasic et al., 2005) (The Math Works, Natick, Massachusetts, USA).

*Sample Entropy (SampEn)*

Sample Entropy (SampEn) was computed according to the procedure published by Richman and Moorman (Richman and Moorman, 2000). Given a finite sequence $x_N = (x_1, x_2,..., x_N)$; we constructed vectors of length $m$, $y_1$ to $y_{N-m}$, defined as

$$y_i = [x_i, x_{i+1},..., x_{i+m-1}], \qquad 1 \leq i \leq N-m \qquad (5)$$

Compute distance between $y_i$ and $y_j$, denoted by $d(y_i, y_j)$, a Chebyshev distance which have to be <r, as

$$d(y_i, y_j) = \max\left(|x_{i+k} - x_{j+k}|\right), \qquad 0 \leq k \leq m-1 \quad j \neq i \qquad (6)$$

For $i = 1,..., N-m$ calculating the probability that any vector $y_j$ that is similar to $y_i$ within $r$ as

$$P_i(m,r) = \frac{n_i(m,r)}{N-m-1} \qquad (7)$$

Where $n_i(m,r)$ is the number of vectors $y_j$ that are similar to $y_i$ subject to the criterion of similarity $d(y_i, y_j) \leq r$. Calculate

$$A(m,r) = \frac{1}{N-m} \sum_{i=1}^{N-m} P_i(m,r) \qquad (8)$$

$$\text{SampEn}(x_N, m, r) = -\ln \frac{A(m+1, r)}{A(m, r)} \qquad (9)$$

SampEn quantifies the irregularity of a time series and estimates the conditional probability that two sequences of m consecutive data points, which are similar to each other (within given tolerance $r$), will remain similar when one consecutive point is included. The SampEn algorithm considers two parameters: tolerance level *r* and pattern length *m*. According to previous studies, we chose a tolerance level of $r = 0.15$ times the standard deviation of the time series and $m = 2$ (Matlab 7.0). Also, the data were analyzed by another SampEn algorithm, in-house written in Java programming language, confirming the initial results.

*Calibration curves*

To elucidate the difference in results between the two nonlinear parameters we used series of surrogate mathematically generated sinusoids with frequencies ranging from 1 to 116 Hz (step 5Hz), while their amplitudes were kept constant. The calibration curves were then analyzed using both algorithms (HFD and SampEn). The breaking point used to construct the calibration curves was defined as $f_b = 0.117 \times f_s$ (Kalauzi et al., 2012). Since the exact position of the calibration curve depends on sinusoid's sampling frequency (Kalauzi et al., 2012), the sampling frequency of the surrogates was set to 1 KHz, to match the actual sEMG sampling rate.

*Statistical Analysis*

The normality of the distribution of all data-sets was examined using Shapiro-Wilk test (Pre and Post-TMS at three levels of contraction). None of the datasets had a normal distribution. Wilcoxon non-parametric rank-sum test was used to compare HFD and SampEn (SPSS Statistical Package for the Social Sciences, Chicago IL release 17.0). $P<0.05$ value was considered statistically significant.

## Results

Figure 1 shows the raw signal recorded at three different contraction levels form FDI muscle.

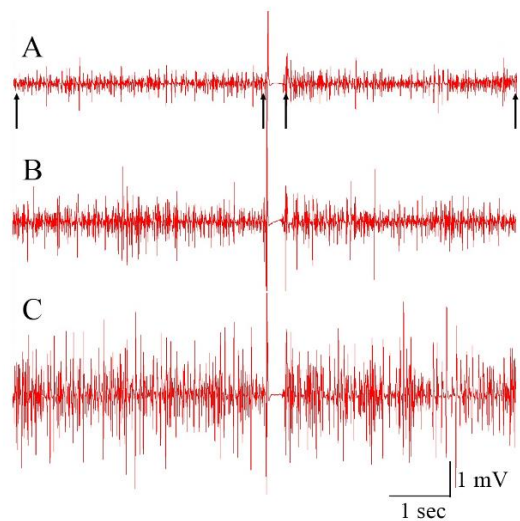

**Figure 1**. Raw sEMG signal at three different contraction levels. Top panel (A) shows mild, middle panel (B) shows medium, and the lower panel (C) shows strong contraction. Arrows indicate the beginning and the end of the segment used for analysis PRE (left side of the recording) and POST TMS (right side of the recording). The HFD of the same segments were: 1.0921/1.0914 (mild), 1.0495/1.047 (medium) and 1.0167/1.015 (strong). The SampEn values of the same segments were: 0.036877/0.035281 (mild), 0.071382/0.056896 (medium) and 0.121227/0.104262 (strong).

Figure 2 and 3 are showing changes in mean SampEn PRE and POST TMS at three different levels of voluntary contraction.

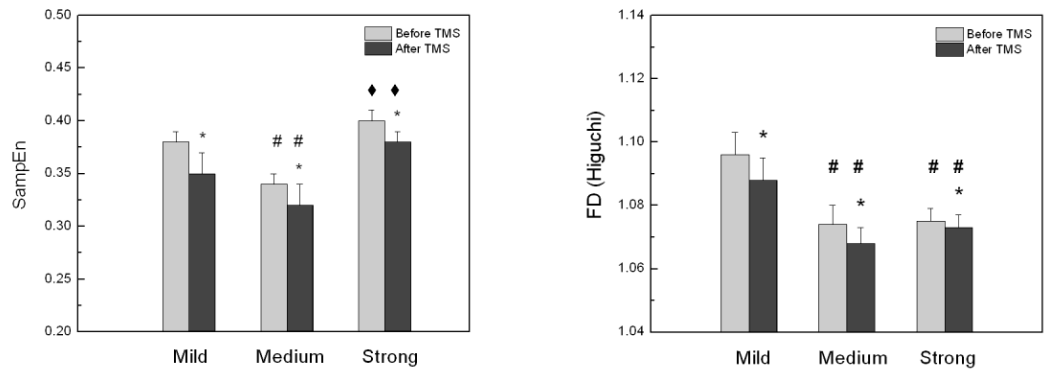

**Figure 2**. (A) Changes in SampEn before and after spTMS. SampEn of sEMG time series before and after spTMS at three levels of contraction. * p< 0.001 PRE vs. POST TMS. # p< 0.001 medium vs. mild contraction, ◆p< 0.001 strong vs. medium contraction. (B) Changes in HFD before and after spTMS. HFD of sEMG time series before and after spTMS at three levels of contraction. * p< 0.001 PRE vs. POST TMS, # p< 0.001 medium and strong vs. mild contraction.

The range of calculated values of SampEn was 0.34 – 0.47 while HFD range was 1.072 – 1.091. SampEn significantly decreased between mild and medium contraction (both PRE and POST comparison) ($p<0.001$), but then significantly increased between medium and strong contraction level ($p<0.001$). There was no significant difference in SampEn between the mild and strong level of contraction ($p>0.05$). SpTMS induced a significant decrease in SampEn (PRE vs. POST comparison, Wilcoxon test) at all levels of muscle contraction ($p<0.001$). The results of Higuchi's Fractal analysis of the same sEMG epochs previously analyzed by SampEn are shown in Figure 2. HFD significantly decreased between mild and medium and mild and strong contraction (both PRE and POST comparison) ($p<0.001$), but not between medium and strong contraction ($p>0.05$). Similarly, to SampEn spTMS induced a significant decrease in HFD POST compared with PRE at all contraction levels ($p<0.001$). Thus, both nonlinear methods show that spTMS induced reduction of the complexity of the signal, irrespective of the contraction intensity while they depicted different changes in complexity between various contraction levels.

To elucidate the difference between HFD and SampEn results of sEMG complexity, and test whether it can be related to their differential sensitivity to the frequency content of the signal, we analyzed mathematically constructed calibration curves (see method), which are shown in Figure 3.

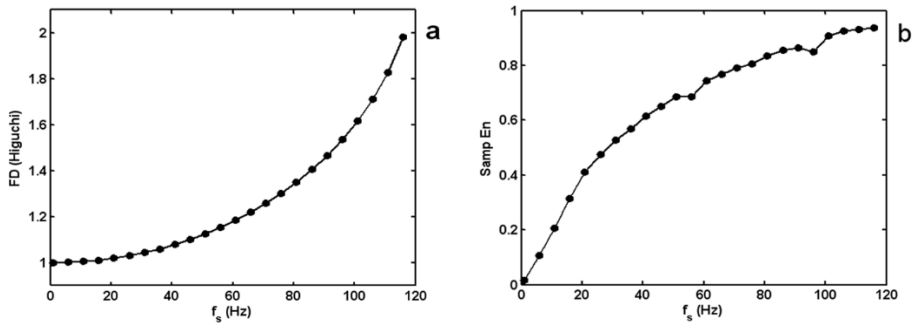

**Figure 3. Comparison of FD and SampEn application on calibration curves.** Calibration curves, showing how Higuchi FD (A) and SampEn (B) depend on frequencies of surrogate sinusoids.

Based on earlier results (Kalauzi et al., 2012) the breaking point for construction of theoretical calibration curves was set to $f_b=0.117 \times f_s$, corresponding to the sampling frequency of 1 KHz. SampEn $(f_i)=\varphi_2(S_i(f_i))$ mapped more linearly in the range $0 < f_i < 60$ Hz (Fig. 4b), whereas HFD$(f_i)=\varphi_1(S_i(f_i))$ mapped the values to more limited range (Fig. 4a). On the other hand, for frequencies $60 < f_i < 120$ Hz SampEn values were mapped to a relatively narrow region (0.8<SampEn<1), while HFD mapped input values to a region occupying approximately 4/5 of the whole HFD range (1.2<HFD<2). This indicates that SampEn and HFD are sensitive to the frequency content of the signal, showing different potential to detect changes in complexity, in a different frequency range (SampEn for lower, HFD for higher frequencies), of a signal under study.

## Discussion

The results of the study confirmed our earlier results that the complexity (HFD) of sEMG decreases with increasing intensity of muscle contraction and is further decreased by spTMS. However, SampEn showed different changes in sEMG complexity, as compared to HFD, while similarly to HFD the complexity was further reduced by spTMS. To elucidate the difference in results we compared the outcome of the analysis of two methods applied on simulated calibration curves and showed that this difference appears to be related to their different sensitivity to the frequency content of the sEMG signal within examined frequency range. The results should be taken into consideration when these nonlinear methods are applied for sEMG analysis.

Our previous study that used HFD showed that the complexity of an sEMG signal significantly decreases with the increase of muscle contraction intensity (Cukic et al., 2013). The results of this study confirm these earlier findings. The reduction of sEMG complexity with increased contraction intensity may appear counterintuitive when compared to changes in sEMG Root Mean Square (RMS) values, characterized by linear relationship between the contraction force and the RMS value. Accordingly, it may be assumed that the increase in force, associated with the increase in muscle unit discharge rate and recruitment, would yield more complex signal. Furthermore, the results of the current study appear at variance to some of the earlier studies, which showed that fractal dimension of the EMG signal rises with the increase of muscle force (Gupta et al., 1997), so that FD could even provide a reasonably good quantification of contraction intensity (Anmuth et al., 1994; Arle and Simon, 1990; Glenny et al., 1991). As argued earlier (Cukic et al., 2013) some of the differences between current and earlier studies may be related to recruitment strategies deployed by CNS when varying muscle force production in different muscles examined in these studies. Namely, it is well-established that muscle force production is largely regulated by motor unit recruitment, (Milner-Brown et al., 1973) with stronger intensities achieved by increased discharge rates (Kukulka and Clamann, 1981; Milner-Brown et al., 1973). In different muscles, the majority of muscle units are recruited at various levels of force production. In biceps brachii 95% of units are recruited at 70% of maximal force production, whereas in FDI muscle, almost all units are recruited at a much lower intensity of around ~ 30% of MVC, with further force increment being generated by frequency modulation (Carpentier et al., 2001; Riley et al., 2008; Staudenmann et al., 2014). Thus, an increase in FD with the rise of voluntary contraction could reach saturation at 70% of MVC (Carpentier et al., 2001; Gitter and Czerniecki, 1995).

Unlike HFD, which decreased between 20% and 40% of MVC and did not further change at 70% MVC, SampEn initially decreased, but then increased between 40% and 70% of the maximal voluntary force. The relationship between muscle force and complexity measured by SampEn was rarely addressed in the past (Cashaback et al., 2013; Zhang et al., 2016). Cashaback et al., (2013) showed that short-term biceps brachii sEMG complexity was moderately influenced by contraction intensity, while the long- term sEMG complexity did not reach statistical significance. On the other hand Zhang et al. (2016) showed strong correlation between SampEn values and the amplitude measurements of the surface EMG signal. At present, it is not clear what may be the reason for this discrepancy although differences between muscles (biceps brachii in Cashaback's study) and subjects (amputees in Zhnag's study) cannot be excluded. It should also be noted that muscle fatigue also decreases the complexity of sEMG (Karthick et al., 2014). Nevertheless, it does not seem likely that it played a role in this study as the contractions were randomized and extra time was availed after each strong contraction to prevent the development of fatigue.

To further examine the difference between HFD and SampEn in relation to muscle contraction intensity we computed the complexity of the signal containing defined frequency spectrum. The analysis of a series of surrogate mathematical sinusoids, $S_i(f_i)$, with monotonously increasing frequencies $f_i = 1, 2, ..., 120$ Hz (Kalauzi et al., 2012) suggested that HFD and SampEn are influenced by the frequency of the underlying signal. SampEn $(f_i)=\varphi_2(S_i(f_i))$ is more linear in the range $0 < f_i < 60$ Hz (Fig. 4b), where HFD$(f_i)=\varphi_1(S_i(f_i))$ tends to map the values to a very limited HFD range (Fig. 4a). On the other hand, for frequencies $60 < f_i < 120$ Hz SampEn values are being mapped into a relatively narrow region (0.8<SampEn<1), while sensitivity of HFD is increased, mapping its input values to a region occupying approximately 4/5 of the whole HFD range (1.2<HFD<2). It is well established that the sEMG frequency spectrum changes with the contraction although the major part of the spectrum are lower than 100Hz, (De Luca, 1984; Knaflitz et al., 1990) other components (harmonics) representing higher frequencies have been described during strong levels of contraction (Christensen et al., 1984; Timmer et al., 1998b). Thus, the difference in detected changes between different intensities of contraction may be related to the differential sensitivity of these two methods to prevailing frequency content of sEMG associated with different contraction intensity. It should be stressed that the connection between fractal dimension and spectral content of the signal, concerning the previous practice of application of solely spectral measures in electrophysiology, was extensively investigated in the past, but none provided the exact mathematical relationship (Weiss et al., 2011). A recent publication by Kalauzi, (Kalauzi et al., 2012) showed the exponential dependence of fractal exponents on the frequency and characterized the relation mathematically. This finding is important as it provides the theoretical framework for the analysis applied in this study. It should also be noted that this allows direct estimation of signal fractal dimension from its Fourier components and establishes that FD does not depend on sinusoid's amplitude and initial phase, but on the frequency of the waveform and sampling frequency.

Finally, the present results confirmed our earlier findings that the complexity of the sEMG signal measured by HFD decreases after cortical spTMS irrespective of the intensity of muscle contraction (Cukic et al., 2013). The present results extend them by demonstrating that an sEMG complexity decreased after spTMS also when estimated by SampEn. As argued earlier the reduction of sEMG complexity after spTMS suggest changes in corticospinal activity, most likely due to transient TMS-induced synchronization of descending excitatory signal (Harris et al., 2008; Marsden et al., 2000; Rosler et al., 2002; Timmer et al., 1998a). Namely, it seems as if spTMS has interupted CNS regulatory mechanisms, making the system less variable and adaptable. It is possible that CNS under these circumstances cannot precisely gauge the state of the corticospinal networks causing the voluntary drive to overshoot, thus causing synchronization, after a sudden externally caused brake of voluntary activity.

Until recently, HFD was thought to be the most sensitive measure for analyzing the complexity of sEMG as suggested by results of analysis that compared the sensitivity of both spectral and nonlinear measures applied on artificially generated EMG (Mesin et al., 2009). However, it has repeatedly been demonstrated that the analysis of complex physiological signals is best performed if different measures/algorithms are used (Ferenets et al., 2006; Kronholm et al., 2007; Stam, 2005) since they may be sensitive to various features of the signal. Thus, the results further reiterate the notion that multiple methods and algorithms should be used to survey the complexity of the signal (Arle and Simon, 1990; Eke et al., 2002; Ravier et al., 2005).

## Conclusions

The results of this study further support increasing body of evidence showing that multiscale approach can quantify subtle information content in physiological time series. They also confirm earlier results that spTMS decreases the complexity of sEMG beyond its immediate electrophysiological effects. Importantly, the data show that SampEn and HFD have different sensitivity in different frequency ranges, making them methodologically complementary for the analysis of sEMG. Finally, based on current results, it could be argued that both methods should be used to elucidate comprehensively changes in complexity of the sEMG signal and thus corticospinal activity. Further studies are needed to explore the duration of spTMS influence on changes in sEMG complexity during voluntary muscle contraction and at different levels of muscle contraction and TMS intensity in healthy and diseased nervous systems. This may provide greater insight into control processes of voluntary control of force in health and disease further expanding the understanding of CNS pathologies.